\documentclass[%
reprint,
superscriptaddress,
amsmath,amssymb,
aps,
]{revtex4-2}

\usepackage{graphicx}
\usepackage{dcolumn}
\usepackage{bm}
\usepackage{siunitx}
\usepackage[rightcaption]{sidecap}
\usepackage{xr}

\makeatletter
\newcommand*{\addFileDependency}[1]{
  \typeout{(#1)}
  \@addtofilelist{#1}
  \IfFileExists{#1}{}{\typeout{No file #1.}}
}
\makeatother

\newcommand*{\myexternaldocument}[1]{%
    \externaldocument{#1}%
    \addFileDependency{#1.tex}%
    \addFileDependency{#1.aux}%
}

\myexternaldocument{SM_new}

\begin{document}

\preprint{APS/123-QED}

\title{Spin-wave bandgap engineering via mode hybridization in dipolar-coupled YIG film/CoFeB nanodisk magnonic crystals}

\author{Junyoung~Hyun}
\affiliation{Department of Applied Physics, Aalto University School of Science, P.O. Box 15100, FI-00076 Aalto, Finland}
\author{Krzysztof~Szulc}
\affiliation{Institute of Molecular Physics, Polish Academy of Sciences, 60-179 Pozna\'n, Poland}
\affiliation{ISQI, Faculty of Physics and Astronomy, Adam Mickiewicz University, 61-614 Pozna\'n, Poland}
\affiliation{CEITEC BUT, Brno University of Technology, 612 00 Brno, Czechia}
\author{Mateusz~Zelent}
\affiliation{ISQI, Faculty of Physics and Astronomy, Adam Mickiewicz University, 61-614 Pozna\'n, Poland}
\affiliation{Fachbereich Physik and Landesforschungszentrum OPTIMAS, Rheinland-Pfälzische Technische Universität Kaiserslautern-Landau, D-67663 Kaiserslautern, Germany}
\author{Nikolai~Kuznetsov}
\affiliation{Department of Applied Physics, Aalto University School of Science, P.O. Box 15100, FI-00076 Aalto, Finland}
\author{Lukáš~Flajšman}
\affiliation{Department of Applied Physics, Aalto University School of Science, P.O. Box 15100, FI-00076 Aalto, Finland}
\author{Maciej~Krawczyk}
\affiliation{ISQI, Faculty of Physics and Astronomy, Adam Mickiewicz University, 61-614 Pozna\'n, Poland}
\author{Paweł~Gruszecki}
\email{gruszecki@amu.edu.pl}
\affiliation{ISQI, Faculty of Physics and Astronomy, Adam Mickiewicz University, 61-614 Pozna\'n, Poland}
\author{Sebastiaan~van~Dijken}
\email{sebastiaan.van.dijken@aalto.fi}
\affiliation{Department of Applied Physics, Aalto University School of Science, P.O. Box 15100, FI-00076 Aalto, Finland}

\date{\today}

\begin{abstract}
We investigate spin-wave transport in hybrid two-dimensional magnonic crystals comprising a low-damping yttrium iron garnet (YIG) film coupled to a periodic array of CoFeB nanodisks. Using propagating spin-wave spectroscopy, super-Nyquist magneto-optical Kerr effect microscopy, and micromagnetic simulations, we demonstrate the formation of pronounced and tunable bandgaps that do not originate from conventional Bragg scattering. Instead, these gaps arise from hybridization between the fundamental magnonic-crystal mode and in-plane transverse standing modes induced by the periodic nanodisk array. The spectral position and width of these gaps are controlled by geometric parameters and by the magnetic state of the nanodisks, including their vortex configuration, which governs both static and dynamic dipolar coupling. For larger lattice periods, additional gaps emerge through hybridization with modes quantized both transverse and parallel to the spin-wave propagation direction, reflecting dispersion folding in two dimensions. Our results establish mode hybridization as a versatile mechanism for engineering spin-wave band structures beyond the constraints of Bragg scattering and provide a pathway toward reconfigurable magnonic devices based on dipolar-coupled hybrid architectures.
\end{abstract}

                              
\maketitle

\section{Introduction}
Magnonics has emerged as a promising platform for wave-based information processing, in which data are encoded in the amplitude and phase of spin waves---collective spin excitations in magnetically ordered materials \cite{Mahmoud_2020,Pirro_2021,Papp_2021,Barman2021,Chumak_2022}. Unlike charge-based electronics, magnonic systems operate without net charge transport, offering the prospect of reduced Joule heating while enabling intrinsically parallel data processing through wave interference and superposition. These features position magnonics as a strong candidate for beyond-CMOS computing and hybrid signal-processing technologies.

A central challenge in this field is the control of spin-wave propagation. A widely used approach relies on engineering the magnetic landscape through periodic modulation of material properties, giving rise to magnonic crystals (MCs)---magnetic metamaterials with periodically modulated parameters \cite{Nikitov2001,Kruglyak2010,Lenk2011,Krawczyk_2014,Chumak2017MC}. Analogous to photonic and phononic crystals, periodic variations in saturation magnetization, effective magnetic field, or geometry lead to Bragg scattering and the formation of band structures with allowed and forbidden frequency ranges. These band structures provide a versatile framework for tailoring spin-wave transport, enabling functionalities such as frequency-selective filtering and phase manipulation.

A broad range of magnonic crystal architectures has been realized, including one-dimensional systems such as grooved films \cite{Chumak2008groove,Qin2018MC}, nanostrip arrays \cite{Gubbiotti2007,Kostylev2008}, and width-modulated waveguides \cite{Chumak2009spin,Frey2020}; two-dimensional antidot lattices  \cite{Duerr2011,Schwarze2012,Levchenko2025} and nanodot arrays \cite{Tacchi2010nanodot,Tacchi2011band}; as well as bicomponent structures \cite{Qin2018MC,Wang2009,Tacchi2012}. In most of these systems, spin-wave transmission is governed by periodic modulation of the effective magnetic field. An important extension is provided by dynamic magnonic crystals, in which the modulation can be tuned via magnetic switching \cite{Topp2010,Szulc_2022}, current-induced Oersted fields \cite{Chumak_2009,Chumak_2010}, voltage control \cite{Wang2017voltage,Merbouche2021}, optical excitation \cite{Vogel_2015,Kuznetsov_2025}, or strain \cite{Sadovnikov2019}, enabling reconfigurable band structures.

Despite this diversity, bandgap formation in most magnonic crystals relies on Bragg scattering along the spin-wave propagation direction \cite{Nikitov2001,Kruglyak2010,Lenk2011,Krawczyk_2014,Chumak2017MC}. In this mechanism, gaps open when the spin-wave wave vector satisfies the Bragg condition imposed by the lattice periodicity, resulting in coherent backscattering at the Brillouin zone boundaries. Although highly effective, this approach intrinsically ties bandgap formation to the propagation direction and lattice periodicity, which can limit flexibility in band-structure engineering, particularly in nanoscale and hybrid systems where additional coupling mechanisms become relevant.

These limitations motivate the exploration of alternative approaches to spin-wave control that exploit mode hybridization, local resonances, or dipolar coupling to nanoscale magnetic elements. Such mechanisms introduce additional degrees of freedom for tailoring spin-wave transmission spectra and enable functionalities that extend beyond the constraints of conventional Bragg scattering.

One such approach is based on magnonic Fabry--Pérot resonators, which offer a compact, interference-based mechanism for controlling spin-wave transport \cite{Qin_2021}. These structures typically consist of a low-loss magnetic film---such as yttrium iron garnet (YIG)---coupled to a patterned ferromagnetic nanostripe and operate in a regime where propagating spin waves remain below the ferromagnetic resonance (FMR) frequency of the nanostripe. Chiral dipolar coupling induces a frequency-downshifted and asymmetric dispersion within the resonator, causing its boundaries to act as interfaces at which the spin-wave wavelength is converted upon reflection and transmission \cite{Talapatra2023}. Multiple reflections give rise to Fabry--Pérot interference, resulting in discrete transmission minima that coexist with low-loss transmission bands. Beyond amplitude control \cite{Qin_2021}, magnonic Fabry--Pérot resonators enable programmable phase shifts via magnetic switching \cite{Lutsenko_2025} and reduce the power threshold for nonlinear transmission \cite{Lutsenko_2026}. 

In this work, we investigate spin-wave transport in two-dimensional magnonic crystals consisting of a continuous, low-loss YIG film coupled to an array of CoFeB nanodisks separated by a thin nonmagnetic spacer. This hybrid architecture combines key features of Fabry--Pérot resonators---specifically dynamic dipolar coupling and field tunability---with those of a two-dimensional magnonic crystal realized in a low-damping YIG film. Both static and dynamic interactions between propagating spin waves in YIG and the periodic nanodisk array can be tuned via geometric parameters, such as disk diameter and lattice period, or reconfigured by modifying the disk magnetization state under an applied magnetic field. Here, we focus specifically on CoFeB nanodisks in the magnetic vortex state. We demonstrate the formation of pronounced gaps in the spin-wave transmission spectra arising from hybridization between the fundamental magnonic-crystal mode and in-plane transverse standing modes. For larger array periods, additional gaps appear through hybridization with modes quantized both transverse to and along the spin-wave propagation direction. 

\begin{figure}
\includegraphics[width=1\linewidth]{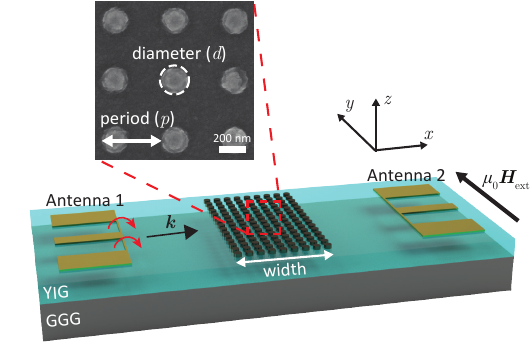}
\caption{\label{fig:fig1} Device schematic and measurement geometry. Spin waves are excited by a microwave antenna and propagate through a 70-nm-thick YIG film integrated with an array of CoFeB nanodisks, before being detected by a second antenna. A nonmagnetic spacer layer (2~nm Ta/4~nm $\mathrm{Ta_2O_5}$) separates the YIG film from the nanodisks, ensuring that the interaction is limited to static and dynamic dipolar coupling. The SEM image above the schematic shows a magnified view of a representative CoFeB nanodisk array with disk diameter $d$ = 180~nm and array period $p$ = 470~nm. Devices with disk diameters $d$ = 120, 180, 240, and 320~nm and array periods $p$ = 390, 470, 550, and 630~nm were fabricated. The array extends over a length of 5~\textmu{}m along the spin-wave propagation direction. An external magnetic field applied along the $y$-axis establishes the Damon--Eshbach transport configuration.}
\end{figure}

\section{Methods}
A 70-nm-thick YIG film was deposited on a (111)-oriented gadolinium gallium garnet (GGG) substrate by pulsed laser deposition (PLD) using a KrF excimer laser ($\lambda$ = 248~nm). Prior to deposition, the substrate was cleaned by sequential sonication in acetone and isopropanol, followed by annealing at 1000~$^{\circ}$C for 4~h in flowing oxygen at 1~bar. The film was deposited at room temperature under an oxygen pressure of 0.025~mbar. The laser operated at a repetition rate of 5~Hz with a fluence of 2.5~J/cm$^2$. After deposition, the sample was annealed at 850~$^{\circ}$C for 4~h in 1~bar of oxygen to promote crystallization of the YIG film.

Arrays of CoFeB nanodisks were fabricated on the YIG film by electron-beam lithography. After development of the poly(methyl methacrylate) (PMMA) resist, a multilayer stack (2~nm Ta/4~nm $\mathrm{Ta_2O_5}$/50~nm CoFeB/3~nm Ta) was deposited by magnetron sputtering, followed by lift-off in acetone. The structural quality of the nanodisks was verified by scanning electron microscopy (SEM; Zeiss Supra 40). Arrays with periods $p$ = 390, 470, 550, and 630~nm and disk diameters $d$ = 120, 180, 240, and 320~nm were fabricated. Microwave antennas were patterned on both sides of the nanodisk arrays using the same lithography process. The antennas consist of 3~nm Ti/120~nm Au, have a width of 500~nm, and are separated by 150~\textmu{}m. In the final device geometry (Fig.~\ref{fig:fig1}), the nanodisk arrays are centered between the antennas and extend 5~\textmu{}m along the spin-wave propagation direction. The arrays are longer than the antennas in the transverse direction to ensure that all excited spin waves pass through the nanodisk region.

Magnetization reversal in the CoFeB nanodisks was characterized by measuring magnetic hysteresis loops using a wide-field magneto-optical Kerr effect (MOKE) microscope (Evico Magnetics) in the longitudinal Kerr configuration. Magnetic fields of up to 200~mT were applied in the plane of the film.

Spin-wave transport between the two microwave antennas was investigated using a home-built propagating spin-wave spectrometer comprising a two-port vector network analyzer (VNA; Agilent N5222A) and a quadrupole electromagnet probe station. Transmission spectra were obtained by recording the $S_{21}$ scattering parameter in frequency-sweep mode. Unless otherwise stated, the external magnetic field was stepped from 60~mT to 0~mT. To enhance contrast, a reference spectrum acquired at 200~mT was subtracted from each measurement. The microwave excitation power was set to $-20$~dBm to ensure operation in the linear spin-wave regime. 

Spin-wave propagation across the CoFeB nanodisk arrays was imaged using super-Nyquist sampling magneto-optical Kerr effect (SNS-MOKE) microscopy. In this technique, a laser frequency comb down-converts GHz magnetization dynamics to an intermediate frequency $\varepsilon$, enabling spin-wave excitation at frequencies $f_{exc}=n\times{f_{rep}}+\varepsilon$. For $\varepsilon \neq 0$, and with the excitation synchronized to the 80~MHz laser repetition rate, the down-conversion is coherent and preserves the phase relationship between the spin waves and the excitation source via lock-in detection at $\varepsilon$. Two-dimensional spin-wave maps were obtained by raster scanning of the sample beneath a stationary laser spot using a piezoelectric stage. Lithographically defined alignment markers on the YIG film ensured consistent laser-spot positioning throughout the measurements.

\begin{figure*}
\includegraphics[width=1\linewidth]{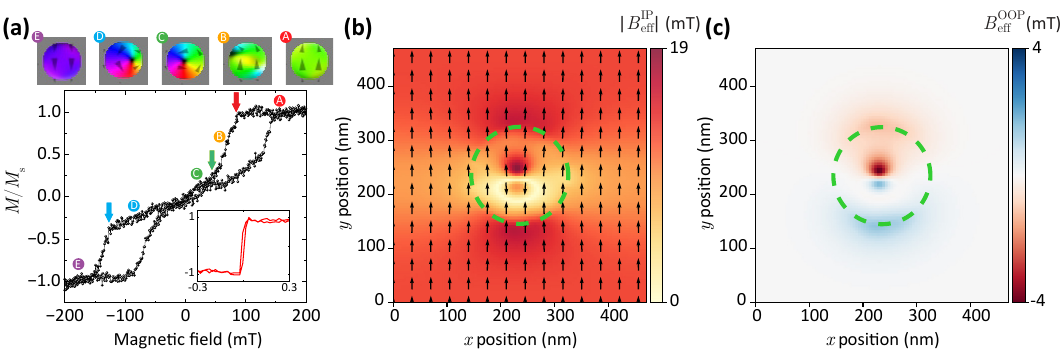}
\caption{\label{fig:fig2} Magnetic switching and effective field distribution in hybrid YIG film/CoFeB nanodisk magnonic crystals. (a)~MOKE hysteresis loop of a CoFeB nanodisk array ($d=180$~nm, $p=470$~nm) fabricated on the 70-nm-thick YIG film. The magnetic contribution of the YIG film is subtracted in the main panel and shown separately in the inset. Micromagnetic simulations at different applied fields illustrate the evolution of the nanodisk magnetization when the field is swept from 200~mT to $-200$~mT. Red, green, and blue arrows indicate the fields corresponding to $S$ state formation, vortex nucleation, and vortex annihilation, respectively. (b)~Simulated in-plane effective magnetic field within the YIG film at an applied field of 12~mT along the $y$-axis. The effective field distribution is averaged over the film thickness. The color scale represents the field magnitude, while arrows denote the field direction. (c)~Simulated out-of-plane effective magnetic field under the same conditions as in (b). Green dashed lines in (b) and (c) outline the lateral position of the CoFeB nanodisk.}
\end{figure*}

Static effective-field distributions and equilibrium magnetization configurations were obtained from micromagnetic simulations performed in MuMax3~\cite{mumax} on a single unit cell with a spatial discretization of $5\times5\times5$~nm$^3$ and periodic boundary conditions applied along both in-plane directions. The applied field was swept from 200~mT to $-200$~mT in 1~mT steps, with the magnetization and effective field recorded at each step.

Field-dependent spin-wave transmission spectra and propagating spin-wave mode profiles were simulated using AMUmax~\cite{amumax}, a fork of MuMax3. The simulation box comprised $13160\times94\times25$ cells of $5\times5\times5$~nm$^3$, including a 70-nm-thick YIG film, a 5-nm-thick nonmagnetic spacer, and fourteen 50-nm-thick CoFeB nanodisks ($d=180$~nm, $p=470$~nm) along the $x$-axis with one structural period along the $y$-axis. Periodic boundary conditions were set to $\mathrm{PBC}_x=8$ and $\mathrm{PBC}_y=256$. At each bias field, all nanodisks were initialized in a vortex state, and the equilibrium configuration was obtained by successive relaxation and energy-minimization steps. Absorbing boundary layers were applied at the $x$ boundaries to suppress spin-wave reflection from the edges of the finite simulation box. Spin waves were excited by a broadband sinc-function field ($b_0=50$~\textmu{}T, $f_\mathrm{cut}=5$~GHz) applied across a 6-\textmu{}m-wide excitation region. Propagating spin-wave maps were reconstructed from the spatial and temporal Fourier transform of the saved $m_x$ component, assembled from individual frequency bins. Further details of the simulation procedure and post-processing pipeline are provided in Section~S1 of the Supplemental Material.

The spin-wave dispersion relations of the hybrid two-dimensional magnonic crystals were obtained by solving the Landau--Lifshitz--Gilbert (LLG) equation using finite-element method (FEM) simulations in COMSOL Multiphysics~\cite{Szulc2025Multifunctional}. The simulation geometry consisted of a single unit cell with Bloch boundary conditions applied along both the $x$ and $y$ directions. The system was first relaxed in a time-domain simulation with the damping increased to $\alpha_{\rm relax} = 0.5$, after which the dispersion relation was extracted using the eigenfrequency solver, with the problem automatically linearized based on the static magnetization configuration obtained in the preceding step.

In all simulations, YIG was modeled using a saturation magnetization $M_{\rm s} = 157.6$~kA/m, an exchange constant $A_{\rm ex} = 3.7$~pJ/m, and a Gilbert damping constant $\alpha = 0.0006$; CoFeB was modeled using $M_{\rm s} = 1150$~kA/m, $A_{\rm ex} = 16$~pJ/m, and $\alpha = 0.005$. The YIG saturation magnetization, film thickness, and damping constant were extracted from FMR spectroscopy and SNS-MOKE microscopy measurements (Fig.~S2 in the Supplemental Material), while the exchange constant was determined by fitting propagating spin-wave spectroscopy spectra to FEM-computed dispersion relations. The material parameters for CoFeB were taken from the literature.

\begin{figure*}
\includegraphics[width=1\linewidth]{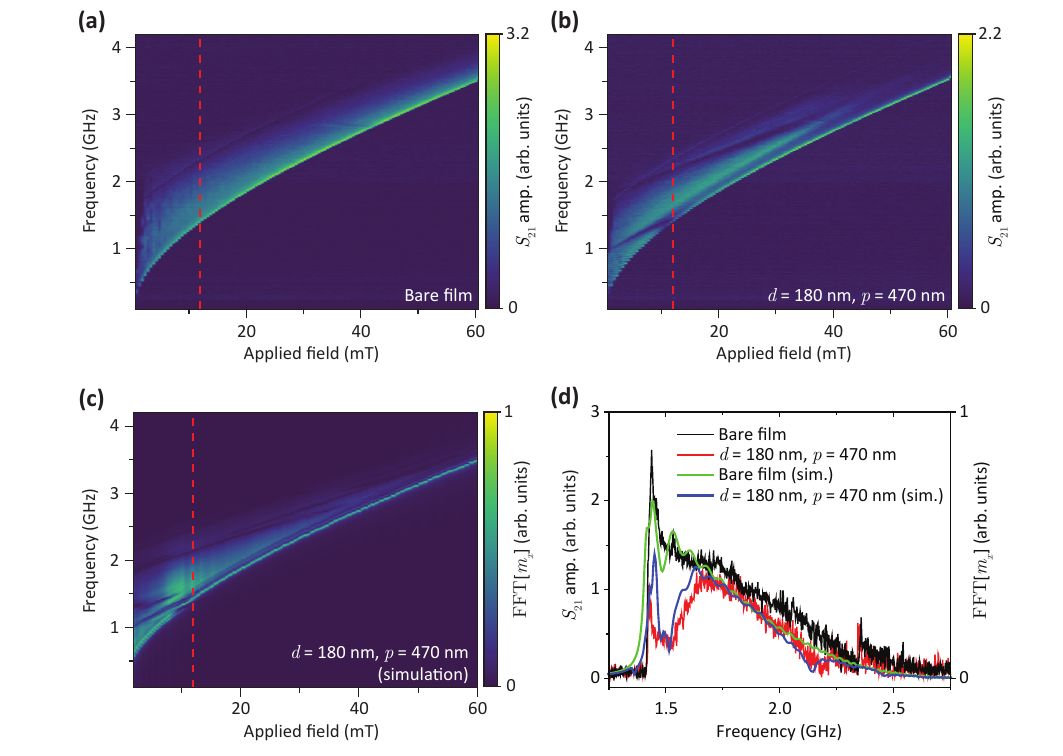}
\caption{\label{fig:fig3} Spin-wave transport in YIG films integrated with CoFeB nanodisk arrays. (a),(b)~Experimental spin-wave transmission maps ($S_{21}$ amplitude) as a function of excitation frequency and applied magnetic field for (a) a bare 70-nm-thick YIG film and (b) the same film with a CoFeB nanodisk array ($d=180$~nm, $p=470$~nm). Experimental spin-wave transmission maps for CoFeB nanodisk arrays with other geometric parameters are provided in Fig.~S4 and S5 of the Supplemental Material. (c)~Simulated spin-wave transmission map for the same hybrid YIG/CoFeB nanodisk system as in (b). (d) Comparison of experimental and simulated spin-wave transmission spectra at an applied field of 12~mT, extracted along the red dashed lines in (a)-(c).}
\end{figure*}

\section{Results and discussion}
\subsection{Spin-wave transmission spectra}
Before presenting the spin-wave transmission spectra, we first characterize the field-dependent magnetic configuration of the hybrid YIG film/CoFeB nanodisk system. Figure~\ref{fig:fig2}(a) shows the MOKE hysteresis loop of a CoFeB nanodisk array ($d=180$~nm, $p=470$~nm), obtained after subtracting the magnetic contribution of the YIG film, which is shown separately in the inset. The nanodisk hysteresis loop exhibits features characteristic of magnetization reversal governed by vortex nucleation and annihilation. For CoFeB nanodisks with a thickness of 50~nm and a diameter of 180~nm, the magnetic ground state is a vortex configuration, in which the in-plane magnetization curls around the disk center to minimize magnetostatic energy while a central core retains out-of-plane magnetization \cite{Shinjo2000}. Under an applied in-plane magnetic field, reversal proceeds sequentially through: (i) formation of an $S$ state---an intermediate configuration in which partial flux closure coexists with field-aligned regions, yielding an overall $S$-shaped magnetization profile; (ii) vortex nucleation at the disk edge; (iii) continuous displacement of the vortex core across the disk; and (iv) vortex annihilation at the opposite edge \cite{Cowburn1999,Guslienko2001}. This sequence, illustrated by the simulated magnetization configurations in Fig.~\ref{fig:fig2}(a), is consistent with the low remanence and nearly reversible hysteresis observed experimentally. From the hysteresis loop, the $S$ state emerges at approximately 80~mT, vortex nucleation occurs near 45~mT, and vortex annihilation takes place at approximately $-130$~mT during a field sweep from 200~mT to $-200$~mT.  

\begin{figure*}
\includegraphics[width=1\linewidth]{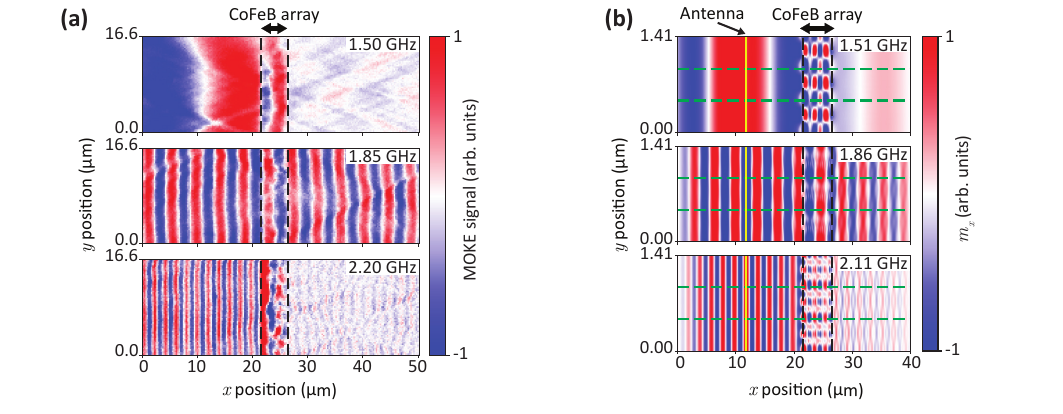}
\caption{\label{fig:fig4} Imaging of spin-wave transport. (a)~SNS-MOKE microscopy maps of propagating spin waves in a 70-nm-thick YIG film with a CoFeB nanodisk array ($d=180$~nm, $p=470$~nm), recorded at excitation frequencies of 1.50, 1.85, and 2.20~GHz under an applied field of 12~mT along the $y$-axis. The selected frequencies correspond to the first transmission gap, the center of the allowed band, and the second transmission gap in the experimental transmission spectrum of Fig.~\ref{fig:fig3}(d). The dashed lines mark the boundaries of the 5-\textmu m-wide nanodisk array. (b)~Corresponding micromagnetic simulation maps at 1.51, 1.86, and 2.11~GHz, showing the real part of the $m_x$ phasor. The spatial extent along $y$ spans only three lattice periods, significantly less than in the experimental maps in (a). The yellow lines indicate the center of the spin-wave excitation region.}
\end{figure*}

The field-driven evolution of the nanodisk magnetization modifies the local effective magnetic field in the YIG film, thereby influencing spin-wave propagation and scattering. Figures~\ref{fig:fig2}(b) and \ref{fig:fig2}(c) show the simulated in-plane and out-of-plane components of the effective magnetic field in the YIG film within a single unit cell of the nanodisk array, calculated at an applied field of 12~mT along the $y$-axis. At this field, the vortex core is displaced from the disk center in the $-x$ direction, consistent with the anticlockwise vortex chirality used in the simulations. The stray field of the out-of-plane magnetized core significantly perturbs both field components in the vicinity of the YIG surface, with rapid spatial decay away from the nanodisks (Fig.~S3 in the Supplemental Material). The field maps in Figs.~\ref{fig:fig2}(b) and \ref{fig:fig2}(c) are averaged over the film thickness. In addition to these static dipolar fields, spin-wave transport in YIG is influenced by dynamic dipolar coupling between the magnetization dynamics in the YIG film and the CoFeB nanodisks. This coupling plays a central role in magnonic Fabry--Pérot resonators formed by a ferromagnetic nanostrip on YIG~\cite{Qin_2021} and is expected to govern the field- and frequency-dependent transmission characteristics of the hybrid YIG film/CoFeB nanodisk system.

We next analyze spin-wave transmission through the YIG film and its modification by the CoFeB nanodisk array. Figures~\ref{fig:fig3}(a) and \ref{fig:fig3}(b) show experimental transmission maps ($S_{21}$ amplitude) as a function of the excitation frequency and applied magnetic field for a bare 70-nm-thick YIG film and for the same film integrated with a CoFeB nanodisk array ($d=180$~nm, $p=470$~nm). Measurements were performed by first saturating the magnetization at 200~mT to acquire a reference spectrum, followed by a stepwise reduction of the field from 60~mT to 0~mT, with $S_{21}$ recorded in frequency-sweep mode at each field step. For the bare YIG film (Fig.~\ref{fig:fig3}(a)), the transmission map is dominated by a pronounced ferromagnetic resonance (FMR) branch. At higher frequencies, the amplitude of spin-wave transmission between the microwave antennas decreases gradually. A clear suppression of transmission occurs near the upper edge of the excitation spectrum (approximately 2.3~GHz at 12~mT), which originates from reduced antenna efficiency in the wavevector range between the first- and second-order excitation modes \cite{Qin_ant_2018}. 

For the YIG film with the CoFeB nanodisk array (Fig.~\ref{fig:fig3}(b)), additional features appear beyond those observed in the bare film. Two transmission gaps emerge: one at low frequencies near the FMR branch and a second at higher frequencies. Both gaps shift to higher frequencies with increasing applied field. In addition, the overall transmission is strongly suppressed above approximately 45~mT, in contrast to the behavior of the bare YIG film. These features arise from the interaction of propagating spin waves with the spatially modulated effective magnetic field induced by the nanodisk array (Fig.~\ref{fig:fig2}) and from dynamic dipolar coupling between the YIG film and CoFeB nanodisks. Consequently, the spin-wave transmission is sensitive to field-induced changes in the nanodisk magnetization configuration. The YIG film magnetization remains nearly uniformly aligned along the $y$-axis throughout most of the investigated field range, with only small local deviations directly beneath the nanodisks at low applied fields. As seen in Fig.~\ref{fig:fig2}(a), the enhancement in transmission near 45~mT in Fig.~\ref{fig:fig3}(b) coincides with the transition of the nanodisk magnetization from the $S$ state to the vortex state when the field decreases. At lower fields, the further evolution of the transmission signal reflects the displacement of the vortex core toward the disk center, which modifies both the effective-field landscape and the strength of dynamic dipolar coupling to propagating spin waves.

Micromagnetic simulations reproduce the main experimental features of of the transmission maps. The simulated map in Fig.~\ref{fig:fig3}(c) can be directly compared to the experimental data in Fig.~\ref{fig:fig3}(b). Both exhibit two transmission gaps with similar field dependencies, confirming that the simulations capture the essential physics of the system. The simulated map also shows additional features at low applied fields that are less pronounced or absent in the experiment. Figure~\ref{fig:fig3}(d) compares the measured and simulated transmission spectra for the bare YIG film and the hybrid system at an applied field of 12~mT. For the YIG/CoFeB nanodisk system, overall agreement is good, with only slight shifts in gap frequencies, likely attributable to small differences between the simulated and actual material parameters, such as the saturation magnetization or exchange stiffness. The first transmission gap is relatively broad, reducing the transmission amplitude by approximately a factor of three compared to the bare YIG film. Transmission within the band between the two gaps remains largely unaffected, indicating weak scattering of spin waves in this frequency range.

Figure~\ref{fig:fig4}(a) shows SNS-MOKE microscopy maps of propagating spin waves in the YIG film with the CoFeB nanodisk array at frequencies corresponding to the first transmission gap (1.50~GHz), the center of the allowed band (1.85~GHz), and the second transmission gap (2.20~GHz) under an applied field of 12~mT. The dashed lines mark the boundaries of the 5-\textmu m-wide nanodisk array. At 1.50~GHz, the spin-wave amplitude is strongly suppressed behind the array region. At 1.85~GHz, spin waves propagate largely unperturbed, with only weak amplitude attenuation and a negligible phase shift. At 2.20~GHz, transmission is again strongly suppressed, and the spin-wave front is noticeably distorted by interactions with the CoFeB nanodisks. These observations are well reproduced by the micromagnetic simulations shown in Fig.~\ref{fig:fig4}(b). The simulation data, plotted on a magnified scale along $y$, reveal transverse mode quantization at frequencies corresponding to the transmission gaps. At 1.51 GHz, the wavevector of the quantized mode along $y$ is $k_y=2\pi/p$, whereas at 2.11 GHz it corresponds to $k_y=4\pi/p$. These features are not resolved in the experimental spin-wave maps due to the limited spatial resolution of the SNS-MOKE technique.

\begin{figure*}
\includegraphics[width=1\linewidth]{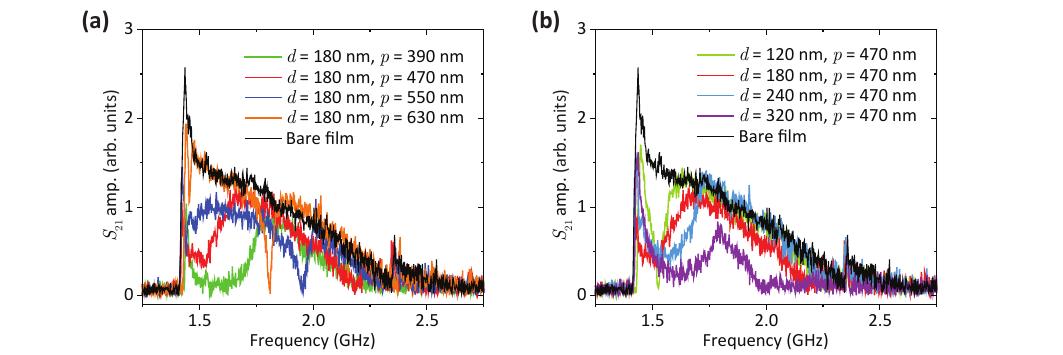}
\caption{\label{fig:fig5} Tuning of spin-wave transport via variation of CoFeB nanodisk array parameters. (a) Experimental spin-wave transmission spectra ($S_{21}$ amplitude) for a fixed disk diameter $d$ = 180~nm and varying array period $p$ = 390, 470, 550, and 630~nm. (b) Spin-wave transmission spectra for a fixed array period $p$ = 470~nm and varying disk diameter $d$ = 120, 180, 240, and 320~nm. All spectra are recorded under an applied magnetic field of 12~mT.}
\end{figure*}

We next examine the dependence of spin-wave transmission on the geometric parameters of the nanodisk array. Figure~\ref{fig:fig5} summarizes spectra recorded at 12~mT for (i) a fixed disk diameter of $d$ = 180~nm with array periods $p$ = 390, 470, 550, and 630~nm, and (ii) a fixed period of $p$ = 470~nm with disk diameters $d$ = 120, 180, 240, and 320~nm. Increasing the array period (Fig.~\ref{fig:fig5}(a)) reduces both the width and depth of the first transmission gap and shifts the second gap to lower frequencies. Increasing the disk diameter primarily broadens the first transmission gap, reflecting enhanced dipolar interaction between the disks and the underlying YIG film. For the largest diameter ($d$ = 320~nm), spin-wave transmission is strongly suppressed across the entire excitation frequency range, consistent with increased spin-wave damping. MOKE measurements confirm that all nanodisks remain in the vortex state at an applied field of 12~mT (Fig.~S6 in the Supplemental Material), indicating that the observed trends are governed predominantly by geometric effects rather than changes in the magnetic configuration of the disks.

\begin{SCfigure*}[0.9][h]
    \includegraphics{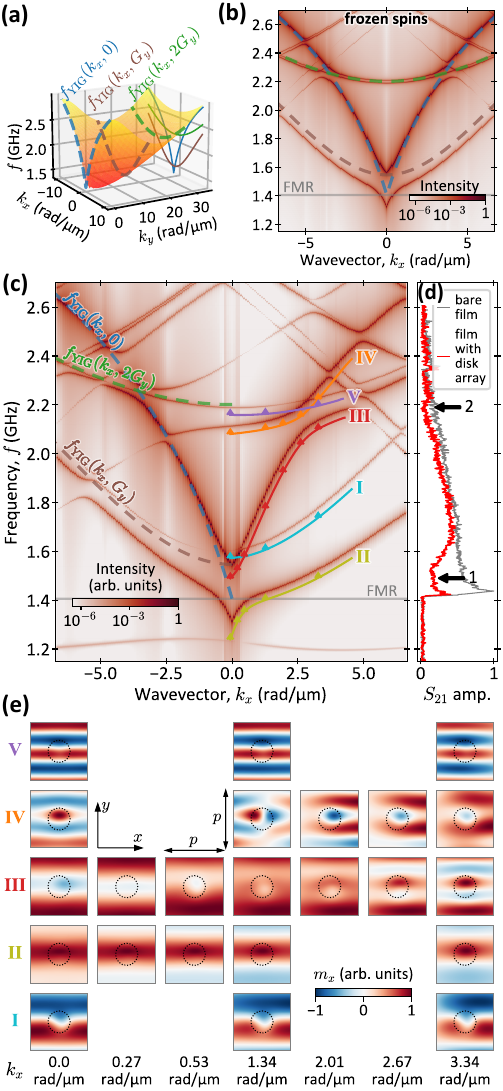}
    \caption{\label{fig:sim-fig2} Numerical analysis of the magnonic crystal band structure for $d$ = 180~nm and $p$ = 470~nm under an applied field of 12 mT. (a)~Two-dimensional dispersion relation of an uncovered YIG film. (b)~Dispersion relation along $k_x$ computed using the frozen-spins method, in which the dynamic magnetization within the CoFeB disks is constrained to zero. The blue, brown, and green dashed lines indicate the bare-film dispersion branches reproduced from (a). (c)~Dispersion relation along $k_x$ incorporating both static and dynamic dipolar coupling between the YIG film and the CoFeB nanodisks. The solid cyan, yellow, red, orange, and purple lines trace dispersion branches I--V (displaced downward from the actual branches for clarity). Upward-pointing triangles mark the $k_x$ values at which mode profiles are shown in panel (e). In (b) and (c), the horizontal gray line indicates the FMR frequency of the uncovered YIG film. (d)~Experimental spin-wave transmission spectrum (amplitude of $S_{21}$) for the uncovered YIG film (gray line) and the hybrid YIG/CoFeB nanodisk magnonic crystal (red line). Arrows mark the first and second transmission gaps. (e)~Spatial mode profiles of the magnonic crystal modes at the ($k_x$, $f$) values indicated in (c), showing the $m_x$ component of the dynamic magnetization in the central plane of the YIG film. Dotted lines indicate the lateral boundaries of the CoFeB disk. Clean versions of (b) and (c) are provided in Fig.~S9 of the Supplemental Material.}
\end{SCfigure*}

\subsection{Analysis of the dispersion relation and mode profiles based on FEM simulations}
To elucidate the physical mechanism underlying the formation of the transmission gaps, we performed FEM simulations. The simulations focused on the spin-wave dynamics arising from the dipolar interaction between the CoFeB nanodisk array and the YIG film, neglecting effects associated with the finite lateral extent of the array along $x$. The system was treated as a two-dimensional magnonic crystal in the limit of a large number of unit cells along the spin-wave propagation direction. Accordingly, the simulations were carried out on a single unit cell of dimensions $p\times p$, with Bloch boundary conditions imposed along the $x$-axis and periodic boundary conditions along the $y$-axis.

The dispersion relation of the reference system---a disk diameter of $d$ = 180~nm and array period of $p$ = 470~nm under an applied field of 12~mT---is presented as a frequency-wavevector colormap in Fig.~\ref{fig:sim-fig2}(c). The plotted intensity at each point in the ($k_x$, $f$) plane is obtained by evaluating the modal intensity
\begin{equation}\label{eq:Imode}
    I_\mathrm{mode}(k_x,f_n(k_x)) = \left| \iiint_{V} m_x(f_n(k_x)) e^{ik_xx}\,\mathrm{d}x\,\mathrm{d}y\,\mathrm{d}z \right|^2,
\end{equation}
where $m_x$ is the $x$ component of the dynamic magnetization, $V$ denotes the volume of magnetic material within a single unit cell, and $f_n(k_x)$ is the complex frequency of the $n$th mode at wavevector $k_x$. The modal intensities are then used to construct a Lorentzian lineshape for each mode, and the contributions from all modes are summed to yield the total spectral intensity
\begin{equation}\label{eq:Ifull}
    I(k_x,f) = \sum_n \frac{I_\mathrm{mode}(k_x,f_n(k_x))}{\mathrm{Im}[f_n(k_x)]\left(1+\left(\frac{f-\mathrm{Re}[f_n(k_x)]}{\mathrm{Im}[f_n(k_x)])}\right)^2\right)}
\end{equation}
at wavevector $k_x$ and frequency $f$.

The resulting band structure is complex, exhibiting pronounced nonreciprocity with respect to $k_x=0$, and comprises multiple bands with distinct intensities, curvatures, and slopes. To facilitate mode identification, we compare the magnonic-crystal dispersion with that of an uncovered YIG film, $f_{\rm YIG}(k_x,k_y)$, shown in Fig.~\ref{fig:sim-fig2}(a).

The fundamental mode of the magnonic crystal derives from the Damon--Eshbach mode $f_{\rm YIG}(k_x,0)$ (dashed blue line in Fig.~\ref{fig:sim-fig2}(c)). The periodicity of the nanodisk array along $y$ additionally gives rise to magnonic-crystal modes associated with bare-film dispersion branches shifted by integer multiples of the reciprocal-lattice vector $G_y=2\pi/p$; the two lowest branches, $f_{\rm YIG}(k_x,G_y)$ and $f_{\rm YIG}(k_x,2G_y)$, are indicated by dashed brown and green lines, respectively. These modes originate from counter-propagating branches at $\pm{nG_y}$: the absence of asymmetry in the dispersion relation along $y$ causes these branches to be degenerate at $k_y=0$, where they hybridize and open a Bragg gap, forming pairs of in-plane transverse standing (IPTS) waves that retain the ability to propagate along $x$ (see Fig.~S7 in the Supplemental Material for the dispersion along $y$). We label the IPTS modes by the integer $n$ corresponding to the multiple of $G_y$ from which they originate. These modes are identified in the full magnonic crystal dispersion of Fig.~\ref{fig:sim-fig2}(c) and are partially traced by the colored solid lines. The remaining bands, with the exception of one, exhibit nearly linear dispersion and large group velocity, consistent with their origin in higher-order branches folded back into the first Brillouin zone. The single exception is a nearly dispersionless mode at approximately 1.2~GHz, which is associated with the gyrotropic resonance of the magnetic vortices within the CoFeB nanodisks.

The two modes comprising each IPTS pair exhibit markedly different characteristics. To analyze their behavior systematically, five branches in Fig.~\ref{fig:sim-fig2}(c) are highlighted with colored solid lines and labeled with Roman numerals I through V. Their spatial mode profiles, characterized by the $m_x$ component of the dynamic magnetization in the central plane of the YIG film, are shown in Fig.~\ref{fig:sim-fig2}(e) for seven values of $k_x$. As a reference, we consider $k_x=1.34$~rad/\textmu{}m (fourth column), where all five branches are well separated in frequency. At this wavevector, mode III displays a nearly uniform magnetization distribution and is identified as the fundamental mode of the magnonic crystal, while pairs I--II and IV--V constitute the first- and second-order IPTS modes, respectively. The mode profiles reveal a systematic distinction within each IPTS pair. The lower-frequency mode of each pair exhibits strong spatial localization of the dynamic magnetization beneath the CoFeB disk, indicative of enhanced coupling to the vortex state. The higher-frequency mode, by contrast, is nearly delocalized, with only a weak reduction in magnetization amplitude in the disk region. 

This spatial localization has a direct consequence for the modal intensity defined in Eq.~(\ref{eq:Imode}). Integration over $y$ and $z$ effectively acts as a projection onto the $k_y=0$ Fourier component of the mode profile. The localized mode, which is predominantly in-phase along $y$, retains a large $k_y=0$ component and therefore yields high modal intensity. The delocalized mode, whose nodal structure along $y$ suppresses the $k_y=0$ component, yields correspondingly low intensity. The modal intensity in turn governs the strength of coupling to the fundamental mode: modes with higher intensity interact more strongly with the nearly uniform mode III. This behavior is directly reflected in both the dispersion relation in Fig.~\ref{fig:sim-fig2}(c) and the mode profiles in Fig.~\ref{fig:sim-fig2}(e). The high-intensity localized modes II and IV couple strongly to mode III, producing large anticrossings with pronounced hybridization of the mode profiles; the anticrossing frequency gaps between modes II and III, and between modes III and IV, are approximately 222~MHz and 108~MHz, respectively. The low-intensity delocalized modes I and V interact weakly with mode III, giving rise to small anticrossings of 14~MHz and 7~MHz, respectively, with mode profiles that remain largely unperturbed across the full wavevector range shown.

To elucidate the origin of mode hybridization, we performed an additional simulation of the same structure using the frozen-spins approach. In this method, the static magnetic configuration is identical to that of the fully coupled simulation, but the dynamic magnetization within the CoFeB disks is constrained to zero ($\mathbf{m}=0$ in the disk volume) during the eigenmode calculation. This allows the contributions of static and dynamic dipolar coupling to be separated: static coupling---arising from the stray field of the disks---is present in both simulations, whereas dynamic dipolar coupling---arising from the precessing magnetization in the CoFeB disks---is active only in the fully coupled case. The resulting frozen-spins dispersion relation is shown in Fig.~\ref{fig:sim-fig2}(b). As in the fully coupled system, the frozen-spins dispersion exhibits a complex band structure in which IPTS modes hybridize with the fundamental mode. Two key differences, however, distinguish the two cases. First, the frozen-spins dispersion is nearly symmetric with respect to $k_x=0$, in contrast to the pronounced nonreciprocity of the fully coupled system (Fig.~\ref{fig:sim-fig2}(c)). Second, the anticrossing gaps are substantially smaller than in the fully coupled case, demonstrating that dynamic dipolar coupling to the vortex state significantly enhances the inter-mode interaction. The frozen-spins dispersion also appears to contain fewer modes. This reduction is, however, not real: only the vortex gyrotropic mode is genuinely absent. The remaining apparently missing modes either nearly overlap due to the restored dispersion symmetry, or fall below the background intensity threshold as a consequence of their mode-profile symmetry. Taken together, these results establish that both the static magnetization configuration and dynamic dipolar coupling to the vortex state contribute to the dispersion relation and, in particular, to the strength of mode hybridization. 

Having established the physical mechanism governing spin-wave propagation in the YIG film region coupled to the CoFeB nanodisks, we compare the FEM dispersion with the experimental transmission spectrum shown in Fig.~\ref{fig:sim-fig2}(d). Two transmission gaps are resolved in the experimental data, marked by arrows labeled 1 and 2. Comparison with the simulated dispersion allows their assignment: transmission gap 1 corresponds to the anticrossing between modes II and III, that is, between the first-order IPTS mode and the fundamental mode, while transmission gap 2 corresponds to the anticrossing between modes III and IV, that is, between the fundamental mode and the second-order IPTS mode. This assignment is confirmed by the quantitative agreement between the calculated anticrossing frequencies and the measured transmission-gap positions.

The field dependence of the transmission gap positions can be understood by approximating the anticrossing frequencies in the coupled system by the crossing frequencies of the bare YIG film dispersion branches: specifically, the DE mode $f_{\rm YIG}(k_x,0)$ with the folded branches $f_{\rm YIG}(k_x,nG_y)$. Within this approximation, the crossing frequencies of these bare-film branches coincide with the observed transmission gap positions across all applied fields examined, confirming that the transmission gaps arise from hybridization between the fundamental mode and the IPTS modes in the region of the YIG film covered by the nanodisk array. The field-dependent character of the first transmission gap is further elucidated by examining the dispersion at 5~mT and 35~mT (Fig.~S8 in the Supplemental Material). At low fields, hybridization between the fundamental mode and the first-order IPTS modes is insufficiently strong to modify the YIG dispersion near $k_x=0$, so the transmission gap remains localized far from the FMR frequency. At high fields, the branch $f_{\rm YIG}(k_x,G_y)$ lies below the FMR of YIG, yet hybridization between the fundamental mode and the first-order IPTS modes persists, sustaining the transmission gap near the FMR frequency.

The same framework accounts for the shift of the second transmission gap with array period observed in Fig.~\ref{fig:fig4}(a). Increasing $p$ reduces $G_y=2\pi/p$, which lowers the frequencies of all $f_{\rm YIG}(k_x,nG_y)$ branches and thereby shifts the IPTS modes---and consequently the second transmission gap---to lower frequencies.

\begin{figure*}
    \includegraphics{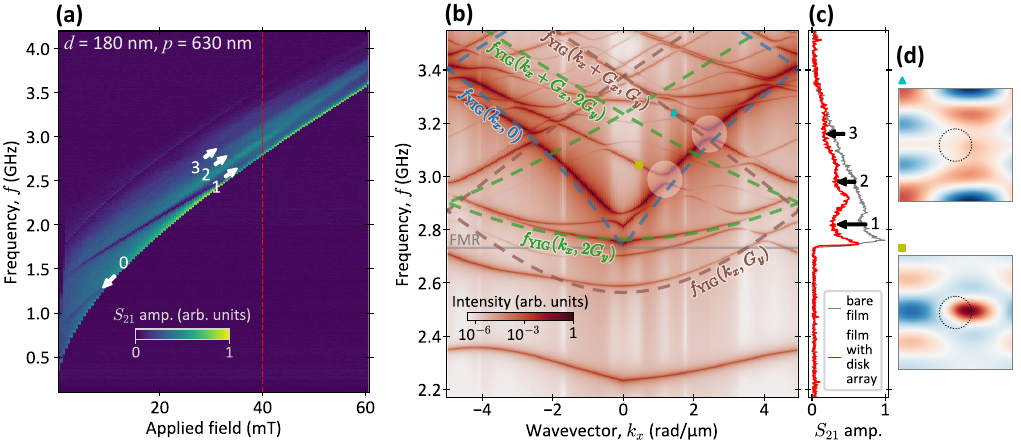}
    \caption{\label{fig:sim-fig6} Analysis of the magnonic crystal
    with $d$ = 180~nm and $p$ = 630~nm. (a)~Experimental spin-wave transmission map ($S_{21}$ amplitude) as a function of excitation frequency and applied magnetic field. (b)~Dispersion relation of the system along $k_x$ calculated for an applied magnetic field of 40~mT. The dashed blue, brown, and green lines indicate the bare-film dispersion branches obtained after backfolding into the first Brillouin zone. The solid gray line marks the FMR frequency of the uncovered YIG film and the white semitransparent circles indicate the hybridizations responsible for the opening of additional transmission gaps. (c)~Experimental spin-wave transmission spectrum (amplitude of $S_{21}$) for the uncovered YIG film (gray line) and the hybrid YIG/CoFeB nanodisk magnonic crystal (red line) at an applied field of 40~mT. In (a) and (c), the arrows with numbers label the transmission gaps. (d) Spatial mode profiles of the magnonic crystal modes at the ($k_x$, $f$) values indicated in (b), showing the $m_x$ component of the dynamic magnetization in the central plane of the YIG film. Dotted lines indicate the lateral boundaries of the CoFeB disk. Clean versions of (b) and (c) are provided in Fig.~S10 of the Supplemental Material.}
\end{figure*}

\subsection{Effect of dispersion folding along the spin-wave propagation direction}
In most magnonic crystals, at most two transmission gaps are observed in spin-wave transmission maps. The hybrid structure with disk diameter $d$ = 180~nm and array period $p$ = 630~nm, whose transmission map is shown in Fig.~\ref{fig:sim-fig6}(a), constitutes an exception: four transmission gaps are observed. The two lowest-frequency gaps (labeled 0 and 1) arise from hybridization between the fundamental mode and IPTS modes associated with dispersion branches shifted by integer multiples of the reciprocal lattice vector along $y$, namely $G_y$ and $2G_y$. The two additional gaps at higher frequencies (labeled 2 and 3) appear only above an applied field of 25~mT. Dispersion analysis shows that these gaps do not originate from fundamental-mode/IPTS-mode hybridization, as the IPTS modes corresponding to $3G_y$ and $4G_y$ lie at significantly higher frequencies.

To elucidate the origin of the additional transmission gaps, we performed FEM simulations for this magnonic crystal at an applied field of 40~mT. The calculated dispersion relation is shown in Fig.~\ref{fig:sim-fig6}(b) and compared to the measured transmission spectrum in Fig.~\ref{fig:sim-fig6}(c) (data extracted along the dashed red line in Fig.~\ref{fig:sim-fig6}(a)). In the calculated dispersion relation, we again superimposed the dispersion branches of the uncovered YIG film, but this time we folded them back into the first Brillouin zone of the magnonic crystal. This procedure increases the number of visible modes compared to Fig.~\ref{fig:sim-fig2}(c). For positive $k_x$ values, two additional anticrossings involving the fundamental mode are identified, marked by semitransparent white circles. The frequencies of these anticrossings coincide with those of transmission gaps 2 and 3, confirming their common origin.

The new anticrossings exhibit contra-directional character, in contrast to the co-directional anticrossings discussed in the previous section. Since all modes in this geometry carry forward character, this indicates that the interacting modes originate from the neighboring Brillouin zone along $k_x$. Comparing the dispersion relation of the coupled system to that of the uncovered YIG film confirms that these modes correspond to branches from the adjacent Brillouin zone. Specifically, the higher-frequency anticrossing (label 3) corresponds to hybridization between the Damon-Eshbach mode $f_{\rm YIG}(k_x,0)$ and the dispersion branch shifted in reciprocal space along both $x$ and $y$, i.e., $f_{\rm YIG}(k_x+G_x,G_y)$. The lower-frequency anticrossing (label 2) corresponds to hybridization between the Damon-Eshbach mode and $f_{\rm YIG}(k_x+G_x,2G_y)$. In this case, the anticrossing gap appears at slightly lower frequency than the bare-mode crossing, which we attribute to mode repulsion between the interacting branches. Spatial mode profiles of the magnonic crystal modes near these high-frequency anticrossings (Fig.~\ref{fig:sim-fig6}(d)) confirm quantization along both $x$ and $y$. 

\section{Conclusion}
In summary, we have demonstrated that coupling a continuous YIG film to a periodic array of CoFeB nanodisks enables a qualitatively different regime of spin-wave control, in which transmission gaps emerge from selective mode hybridization rather than conventional Bragg reflection. Combining experiment and simulation, we have identified the microscopic origin of these gaps as anticrossings between the fundamental magnonic crystal mode and a hierarchy of in-plane transverse standing modes, with coupling strengths governed by the spatial localization of dynamic magnetization. The resulting band structure can be continuously tailored through lattice geometry and the magnetic-field-controlled states of the nanodisk, providing access to multiple, independently tunable gaps and complex dispersion features arising from two-dimensional periodicity. The vortex configuration of the nanodisks plays a central role in this tunability: its field-dependent stray field and dynamic dipolar coupling to the YIG film directly control the hybridization strength and gap frequencies. These findings establish hybrid magnetic film/nanodisk systems as a versatile platform for magnonic band-structure engineering and suggest new routes toward compact, low-loss, and reconfigurable spin-wave functionalities based on controlled mode coupling in planar architectures.

\begin{acknowledgments}
This project has received funding from the European Union’s Horizon Europe research and innovation program under Grant Agreement No. 101070347-MANNGA. However, views and opinions expressed are those of the authors only and do not necessarily reflect those of the European Union or the European Health and Digital Executive Agency (HADEA). Neither the European Union nor the granting authority can be held responsible for them. The project also received funding from the Jane and Aatos Erkko Foundation and the Technology Industries of Finland Centennial Foundation through the Future Makers funding program, the Research Council of Finland (Project No. 357211), and the National Science Center of Poland (OPUS-LAP Grant No. 2020/39/I/ST3/02413). We acknowledge the provision of facilities by Aalto University at the OtaNano-Nanomicroscopy Center (Aalto-NMC) and the computational resources provided by the Aalto Science-IT project.
\end{acknowledgments}

\section*{Data Availability}
The data that support the findings of this article are openly available at [].

\bibliography{main_ref}

\end{document}